\documentclass[aps,twocolumn,nofootinbib,showpacs,prd]{revtex4}
\usepackage{graphicx}
\usepackage{amsfonts}
\usepackage{amssymb}
\usepackage{amsbsy}
\usepackage{amsmath}
\usepackage{latexsym}
\usepackage{natbib}
\usepackage{bm}
\usepackage{color}
\usepackage{float}
\usepackage[utf8]{inputenc}

\begin{document}
\title{Phase Transition and Critical Behavior in Gravitational Collapse}
\author{Soumya Chakrabarti\footnote{soumya.chakrabarti@vit.ac.in}}
\affiliation{School of Advanced Sciences, Vellore Institute of Technology, \\ 
Tiruvalam Rd, Katpadi, Vellore, Tamil Nadu 632014 \\ India}


\date{\today}

\begin{abstract}
We present a thermodynamic analysis of spherically symmetric gravitational collapse. Using the Hayward-Kodama formalism, we treat a collapsing sphere as a thermodynamic system and express the surface gravity $\kappa_{hk}$ in terms of the geometric variables. We derive the specific heat capacities and identify a critical condition $\dot{\kappa}_{hk} = 0$ as the locus of second order phase transition during the collapse. Through specific examples, we demonstrate that the condition is independent of singular/non-singular nature of the geometry. We also find that the critical condition of phase transition is equivalent to a stationary condition of the expansion of null congruences. This establishes a direct correspondence between geometric stability and thermodynamic criticality, allowing the identification of apparent horizon as a universal critical surface in the phase-space of gravitational collapse.
\end{abstract}

\maketitle

\section{Introduction}
The study of gravitational collapse remains one of the most important and enigmatic problems in General Relativity (GR). The history of the subject spans almost a century, beginning from the pioneering work of Oppenheimer and Snyder \cite{oppenheimer_snyder}, eventually evolving into modern numerical and semi-analytical approaches \cite{joshi_review, collapse_review}. The geometric end-states of a gravitational collapse have always provided an essential test-bed for a relativist to interpret concepts like the event horizon to interpret singularity theorems or the cosmic censorship. It is usually accepted that a gravitational contraction leads either to the formation of a black hole, where the singularity is hidden behind an event horizon, or to a naked singularity, visible to distant observers. These two distinct outcomes provide the primary motivation of the so-called Cosmic Censorship Conjecture, which in principle forbids an occurrence of observable singularities \cite{penrose1, penrose2, goncalves, christ1, christ2, gold}.  \medskip

In parallel, the discovery of black-hole thermodynamics and its subsequent generalizations reveal that horizons can be considered as thermodynamic entities, characterized by temperature, entropy and associated energy flux \cite{bekenstein, hawking, padmanabhan_thermo}. From this viewpoint, the emergence of horizon during a gravitational collapse should signal transitions between two different gravitational phases under the influence of strong gravity \cite{cai_thermo}. Such a thermodynamic interpretation of spacetime geometry has a long and intricate history. The foundational ideas of Bekenstein and Hawking \cite{bekenstein, hawking} led to the notion that stationary black holes possess an entropy proportional to their horizon area and radiate thermally at a temperature determined by their surface gravity. Jacobson's seminal result \cite{Jacobson1995} further elevated this correspondence by deriving the Einstein field equations from Clausius relation $\delta Q = T\,dS$, assuming the proportionality of entropy to the area on local Rindler horizon. Hayward introduced the concept of a unified first law \cite{Hayward1998}, identifying a dynamical surface gravity $\kappa_{hk}$ defined with respect to the Kodama vector field \cite{kodama}, associated to any spherically symmetric trapping horizon. This formalism permits the definition of an effective temperature $T_h = |\kappa_{hk}|/(2\pi)$ and horizon entropy proportional to the instantaneous area, thereby generalizing black-hole thermodynamics into dynamical geometries \cite{Akbar1, Akbar2}. \medskip

For a dynamical spacetime going through gravitational collapse, the relevant boundary surface separating the trapped interior from the untrapped exterior is called an \emph{apparent horizon} \cite{Hawking1973, Hayward1994}. It is a time-evolving hypersurface characterized by the vanishing of $g^{\mu\nu} \, Y_{,\mu} \, Y_{,\nu}$, where $Y(r,t)$ denotes the areal radius in spherical symmetry. The surface gravity and temperature of such an apparent horizon can be defined via the Hayward-Kodama prescription \cite{kodama, Hayward1998, Hayward1998b} and this association has been used to extend the laws of black-hole thermodynamics into simple, homogeneous, non-stationary configurations such as cosmological models \cite{Bak2000, Cai2005, Akbar1, Akbar2}. Moreover, recent developments in horizon thermodynamics have also revealed that the heat capacities associated with dynamical horizons may exhibit divergence during deceleration-to-acceleration transition of the universe \cite{Padmanabhan2002, Paranjape2006, Banerjee2008, Banerjee2010, booth, Chakrabarti2024}. However, the possible occurrence of phase transitions during a gravitational collapse has remained unexplored. In this article, we show that the apparent horizon can indeed be considered as a critical surface, separating two phases of the collapsing system characterized by focussing of geodesics. We first revisit a general thermodynamic description of spherically symmetric gravitational collapse. Starting from the Einstein field equations with an effective perfect-fluid source, we identify the apparent horizon and find the corresponding Hayward-Kodama temperature. By computing the rate of change of total entropy inside the horizon, we derive the specific heat capacities and explore their divergence, if any. This allows us to correlate phase transitions and the underlying criticality of a collapse in analogy with standard thermodynamics. \medskip

We also find that the condition $\dot{\kappa}_{hk}=0$ corresponds to a stationary point of the expansion of outgoing null geodesics, as governed by the Raychaudhuri equation. At the thermodynamic critical point, the net focusing term momentarily vanishes, signaling a transient equilibrium in the geometric flow. Thus, the critical condition can be interpreted as a thermodynamic analogue of the geodesic focusing extremum, linking the stability of the apparent horizon to the kinematic stability of null congruences. Under suitable approximation near the critical condition we also interpret this phase transition thermodynamically, by identifying a critical exponent and an order parameter, in analogy with spontaneous magnetization in a ferromagnetic transition or the condensate amplitude in a Landau-Ginzburg model.  \medskip

The remainder of the paper is organized as follows. In Section $II$ we briefly revisit the formalism to define a Hayward-Kodama surface gravity and associated temperature. We point out, methodically, the way forward to derive the specific heat capacities procedd to explore three distinct examples : the Lemaitre-Tolman-Bondi collapse in Section $III$, the conformally evolving Joshi-Malafarina-Narayan naked singulariy in Section $IV$ and the conformally evolving Simpson-Visser metric in Section $IV$. In Section $V$, we identify the critical condition governing the locus of a second order phase transition during gravitational collapse. Section $VI$ includes a mathematical analogy between the critical condition and the behavior of null congruence leading into a stationary condition of null expansion. In Section $VII$, we try to identify the critical exponent and the order parameter associated with this phase transition during collapse, under suitable approximations. Similarly, in Sections $VIII$ and $IX$, we discuss approximate behavior of the specific heat capacities in a near-horizon and a near-singularity limit respectively. We re-interpret the nature of the critical condition in these conditions and summarize our findings in Section $X$.   

\section{Apparent horizon, Kodama-Hayward surface gravity and The First law of Thermodynamics}
We initiate the formalism with a generic spherically symmetric metric
\begin{equation}\label{metric}
ds^2 = -A^2(t,r) dt^2 + B^2(t,r) dr^2 + C^2(t,r) d\Omega^2.
\end{equation}
We adopt geometric units $G = c = 1$. The Einstein equations, $G_{\mu\nu} = 8\pi T_{\mu\nu}$, for an effective perfect fluid in a comoving frame translates into
\begin{equation}\label{rho_p_def}
\rho = -\frac{1}{8\pi}G^t{}_t,\qquad p = \frac{1}{8\pi}G^r{}_r.
\end{equation}

To calculate the apparent horizon and the associated surface gravity we need the two-dimensional normal metric written as $h_{ab}dx^a dx^b = -A^2 dt^2 + B^2 dr^2$. The apparent horizon is defined as the outer marginally trapped surface and determined by the condition
\begin{equation}\label{AHcond}
h^{ab}\partial_a C\,\partial_b C = 0,
\quad\Longleftrightarrow\quad
-\frac{(\partial_t C)^2}{A^2} + \frac{(\partial_r C)^2}{B^2} = 0.
\end{equation}
The Kodama vector is defined on the two-dimensional normal metric by $K^a \equiv \epsilon^{ab}\nabla_b C$, where $\epsilon^{ab}$ is the antisymmetric volume form. Using the Kodama vector one can derive an expression for the Hayward-Kodama surface gravity as
\begin{equation}\label{kappa_box}
\kappa_{hk} \;=\; \frac{1}{2}\,\Box C
\;=\; \frac{1}{2\sqrt{-h}}\partial_a\!\Big(\sqrt{-h}\,h^{ab}\partial_b C\Big).
\end{equation}
For the two-dimensional metric, we use $h^{tt} = -A^{-2}$, $h^{rr}=B^{-2}$, $\sqrt{-h}=AB$ and expand Eq. \eqref{kappa_box} into an explicit form
\begin{equation}\label{kappa_explicit}
\kappa_{hk} \;=\; \frac{1}{2AB}\left[
-\partial_t\!\Big(\tfrac{B}{A}\,\partial_t C\Big)
+\partial_r\!\Big(\tfrac{A}{B}\,\partial_r C\Big)
\right].
\end{equation}

Using $\kappa_{hk}$ one can also define the Hayward-Kodama temperature of the apparent horizon as
\begin{equation}\label{Th_def}
T_h \;=\; \frac{|\kappa_{hk}|}{2\pi}.
\end{equation}

We imagine the collapsing distribution as a sphere with area $A_h = 4\pi C_h^2$ where $C_h$ is the areal radius. Using this, we define the horizon entropy as
\begin{equation}\label{Sh_def}
S_h = 2\pi~,~ A_h = 8\pi^2 C_h^2 ~,~\dot S_h = 16\pi^2 C_h \dot{C_h},
\end{equation}
where the overdot denotes a time derivative. Naturally, the volume enclosed by the horizon can also be calculated as
\begin{equation}\label{V_def}
V = \frac{4\pi}{3}C_h^3 ~,~ \dot V = 4\pi C_h^2 \dot{C_h}.
\end{equation}

We adopt the first law of thermodynamics and write it in a simplified form
\begin{equation}\label{firstlaw}
T_h dS_{in} = dU + p dV ~~,~~ U=\rho V.
\end{equation}

Using Eq. \eqref{V_def} we write the explicit rate of change of entropy as
\begin{equation}\label{Sdot_general}
\dot{S_{in}} = \frac{1}{T_h}\Big[ (\rho+p) 4\pi C_h^2 \dot{C_h} + \frac{4\pi}{3}{C_h}^3  \dot{\rho} \Big].
\end{equation}
Using Eq. \eqref{rho_p_def} we can also express Eq. (\ref{Sdot_general}) in terms of the Einstein tensor as
\begin{equation}\label{Sdot_geometry}
\dot{S_{in}} = \frac{1}{2T_h}\big(-G^{t}_{t} + G^{r}_{r}\big) C_h^2 \dot{C_h} - \frac{C_h^3}{6T_h} \partial_t G^{t}_{t}.
\end{equation}

In order to calculate the specific heat capacities of the collapsing system we need only the first order change in entropy as in Eq. \eqref{Sdot_geometry}. We retain the standard thermodynamic definitions as
\begin{equation}\label{CvCp_beta_def2}
C_V = T_h\left(\frac{\partial S_{in}}{\partial T_h}\right)_{V} ~~,~~ C_P = T_h\left(\frac{\partial S_{in}}{\partial T_h}\right)_{P}.
\end{equation}

As explored in literature, in a spatially flat, homogeneous and isotropic universe, these specific heat capacities show divergence whenever there is a deceleration to acceleration transition of the expanding universe, signalling a second order phase transition. Our motivation here is to explore and identify such a divergence during gravitational collapse and pose it as a generic property of the underlying space-time geometry. To that end we choose, three particular examples, all of which can provide different probable end-states of a collapse.
 
\section{Thermodynamic Analysis for a Collapsing Lemaitre-Tolman-Bondi Spacetime}
An inhomogeneous spherical collapse that produces a black hole is well described by the Lemaitre-Tolman-Bondi (LTB) metric \cite{lem, tol, bond}, written as
\begin{equation}
ds^2 = -dt^2 + \frac{R'^2}{1-f(r)}dr^2 + R^2 d\Omega^2,
\end{equation}
for which the Misner-Sharp mass function is defined as $F(t,r) = R(\dot{R}^2 + f)$. It is a convenient choice to write the Einstein equations for this metric in terms of the mass function, as
\begin{equation}\label{feltb}
8\pi\rho = \frac{F'}{{R}^2 R'}~~,~~ 8\pi p = -\frac{\dot{F}}{R^2 \dot{R}}.
\end{equation}

We derive the Hayward-Kodama surface gravity as
\begin{equation}
\kappa_{hk} = -\frac{1}{2R'}\left( R'\ddot{R} + \dot{R} \dot{R'} + \frac{1}{2} f'(r) \right) \equiv -\frac{N}{2R'},
\end{equation}
where we have defined, for convenience
\begin{equation}
N =  R'\ddot{R} + \dot{R} \dot{R'} + \frac{1}{2} f'(r).
\end{equation}
Naturally, the Kodama temperature is defined as
\begin{equation}
T = \frac{\kappa_{hk}}{2\pi} \longrightarrow \dot T = \frac{\dot{\kappa_{hk}}}{2\pi} = -\frac{R'\dot{N} - N \dot{R'}}{4\pi R'^2}.
\end{equation}

The areal volume and its time derivative are $V = \frac{4\pi}{3}R^3$ and $\dot{V} = 4\pi R^2 \dot{R}$. Considering the first law of thermodynamics, the specific heat at constant volume becomes 
\begin{eqnarray}
C_V &=& V \frac{d\rho}{dT}\\
 &=& -\frac{2\pi R^2}{3}\Bigg\lbrace \frac{-3\dot{F} R'^2 + \dot{F'} R R' - F'R\dot{R'} + F'\dot{R}R'}
{R'\dot{N} - N\dot{R'}}\Bigg\rbrace.
\label{CVfinal}
\end{eqnarray}
Similarly, we define
\begin{equation}
C_P = V\frac{d\rho}{dT} + (\rho + p)\frac{dV}{dT},
\end{equation}
and derive the explicit form as 
\begin{eqnarray}\nonumber
&& C_P = -\frac{2\pi R^2}{3(R'\dot{N} - N\dot{R'})} \Big\lbrace -3\dot{F} R'^2 + \dot{F'} R R' \\&&
- F'R\dot{R'} + F'\dot{R} R' + 3\Big(\frac{F'}{R'^2} - \frac{\dot{F}}{\dot{R} R'}\Big)R'\dot{R}\Big \rbrace.
\label{CPfinal}
\end{eqnarray}

Eqs. (\ref{CVfinal}) and (\ref{CPfinal}) provide an understanding of the underlying thermodynamic nature of the collapse, through the denominators in the expressions of the specific heat capacities. It is straightforward to note that whenever $(R'\dot{N} - N\dot{R'}) = 0$, the heat capacities diverge, signalling a second order phase transition. One can also represent the denominator as the first rate of change of the horizon temperature or the surface gravity, i.e., $\dot{\kappa_{hk}}$.

\section{Thermodynamic Behaviour of the Joshi-Malafarina-Narayan Naked Singularity Metric with a Time-Evolving Conformal Factor}
As we have mentioned in the introduction, a black hole is not the only possible candidate to be borne out of a gravitational collapse. We now extend the analysis to a second candidate, a naked singularity. We choose the Joshi-Malafarina-Narayan (JMN) metric \cite{Joshi2011, rajibul} but modify it slightly by including an evolving conformal factor. The static geometry can model the final static configuration resulting from a gravitational collapse that does not form an event horizon. Incorporating the conformal factor therefore allows us to probe how local thermodynamic quantities evolve near the singular core. The static JMN metric is given by

\begin{equation}
\label{eq:JMN_static}
ds^2 = -\left(1 - M_0\right)\left(\frac{r}{r_b}\right)^{\frac{M_0}{1-M_0}} dt^2 + \frac{dr^2}{1 - M_0} + r^2\, d\Omega^2,
\end{equation}
where $0 < M_0 < 1$ is a constant and $r_b$ is the matching radius to a Schwarzschild exterior. The singularity at $r=0$ is globally visible. We introduce a conformal factor $\Phi(t,r)$, and write
\begin{eqnarray}\label{eq:JMN_conf}
&& ds^{2} = \Phi^{2}(t,r)\Big[-f(r)\,dt^{2} + \frac{dr^{2}}{g(r)} + r^{2}d\Omega^{2}\Big], \\&&
f(r)=(1-M_{0})\Big(\frac{r}{r_{b}}\Big)^{\frac{M_{0}}{1-M_{0}}}~~,~~ g(r)=1-M_{0}.
\end{eqnarray}
The two dimensional normal metric is written as
\begin{equation}
h_{ab}dx^{a}dx^{b} = \Phi^{2}(t,r)\Big[-f(r)\,dt^{2}+\frac{dr^{2}}{g(r)}\Big].
\end{equation}
The area-radius is $C(t,r)=\Phi(t,r) r$, and we use it to derive the Hayward-Kodama surface gravity as
\begin{equation}
\kappa_{hk} = \frac{1}{2\Phi^{2}}\Big[ -\frac{r \ddot{\Phi}}{f} + \frac{1}{\sqrt{fg}}\partial_{r}\big\lbrace\sqrt{fg} (\Phi + r\frac{\partial \Phi}{\partial r})\big\rbrace\Big].
\end{equation}

The hayward-Kodama temperature is, as usual,
\begin{equation}
T(t,r)=\frac{|\kappa_{hk}|}{2\pi}.
\end{equation}

The volume enclosed within a sphere of areal radius $C(t,r)=\Phi(t,r) r$ is derived as
\begin{equation}
V = \frac{4\pi}{3}C^{3} = \frac{4\pi}{3}\Phi^{3} r^{3}.
\end{equation}

Using the definitions of heat capacities from the last section,
\begin{equation}
C_{V} = V\,\frac{d\rho}{dT}~~,~~C_{P}(t,r) = V \frac{d\rho}{dT} + (\rho+p) \frac{dV}{dT}.
\end{equation}

and the Einstein field equations, we derivative the exact forms as

\begin{eqnarray}
&& C_{V} = \frac{8\pi^{2}\Phi^{3}r^{3} \dot{\rho}}
{\dot{\kappa_{hk}}}, \\&&
C_{P} = \frac{8\pi^{2}\Phi^{3}r^{3} \dot{\rho}}
{\dot{\kappa_{hk}}} + (\rho+p) \frac{8\pi^{2}\Phi^{2}r^{3}\Phi_{t}}{\dot{\kappa_{hk}}}.
\end{eqnarray}

\section{Thermodynamic Behavior of The Conformally Scaled Simpson-Visser metric}
\label{sec:SV_conf_thermo_flipped}
We next consider a collapsing configuration that can lead to a non-singular final state, described by a Simpson-Visser metric \cite{simpsonvisser}. This geometry modifies the standard Schwarzschild spacetime by introducing an additional parameter that smoothly interpolates between a classical black hole and a Morris-Thorne type traversable wormhole. Recent studies \cite{SCSK} have shown that a collapsing system can dynamically evolve into a generalized Simpson-Visser configuration. Earlier, Roman and Bergmann \cite{roman} constructed similar models of singularity-free spherical collapse involving weak energy condition violations. The motivation for such regular end-states stems from the broader class of non-singular black holes \cite{Bardeen:1968, Bergmann-Roman, Hayward:2005}. The standard Simpson-Visser line element is given by
\begin{eqnarray}\nonumber \label{simpsonvisser}
&&ds^{2} = -\Big(1-\frac{2m}{\sqrt{l^{2}+a^{2}}}\Big)dt^{2} + \frac{dl^{2}}{1-\frac{2m}{\sqrt{l^{2}+a^{2}}}} \\&& + \Big(l^{2}+a^{2}\Big)\Big(d\theta^{2}+\sin^{2}\theta d\phi^{2}\Big),
\end{eqnarray}
where the coordinates vary over 
$l\in(-\infty,+\infty)$ and $t\in(-\infty,+\infty)$. The parameter $a$ determines the geometry: for $a=0$, the Schwarzschild spacetime is recovered; $a>0$ corresponds to either a regular black hole or a traversable wormhole configuration. No curvature singularity appears, since all curvature invariants remain finite at $l=0$. Defining $l^{2}+a^{2}=r^{2}$ we transform the metric in Eq. (\ref{simpsonvisser}) into
\begin{equation}\label{beginning}
ds^{2} = -\Big(1-\frac{2m}{r}\Big)dt^{2} + \frac{dr^{2}}{\Big(1 - \frac{a^{2}}{r^{2}}\Big)\Big(1-\frac{2m}{r}\Big)} + r^{2} d\Omega^2,
\end{equation}
with $r\in(a,+\infty)$. Moreover, in order to describe dynamical collapse, we extend this into a time-dependent form,
\begin{equation} \label{beginning2}
ds^{2} = -A(r,t)^{2} dt^{2} + \frac{dr^{2}}{\Big(1 - \frac{b(r)}{r}\Big)B(r,t)^2} + C(r,t)^{2} d\Omega^2,
\end{equation}
where
\begin{eqnarray}\label{coordinate}\nonumber
&&r\in(r_{w},+\infty);\\&& 
t\in(0,+\infty),
\end{eqnarray}
and $r_{w}$ denotes the wormhole throat, equivalent to the parameter $a$. We consider the conformally scaled spherically symmetric metric
\begin{equation}\label{eq:conf_SV_flipped}
ds^{2} = T(t)^{2}\Big[-A(r)^{2}dt^{2} + B(r)^{2}dr^{2} + r^{2}d\Omega^{2}\Big],
\end{equation}
where \(A(r)\) and \(B(r)\) are metric functions, and
\begin{equation}\label{eq:B_def_flipped}
B(r)=\frac{1}{\big(1-\frac{b(r)}{r}\big)^{1/2}A(r)}.
\end{equation}
Here \(T(t)\) is a positive, time-dependent conformal factor. The areal radius of the geometry is $R(t,r) = T(t) r$. We derive the Hayward-Kodama surface gravity as
\begin{equation}\label{eq:kappa_SV_conf}
\kappa = -\frac{r \ddot{T}}{2T^{2}A^{2}}
+\frac{1}{2 T AB}\partial_{r} \Big\lbrace\frac{A}{B}\Big\rbrace.
\end{equation}

For a sphere of areal radius $R(t,r) = T(t) r$, the volume and its derivative is calculated as
\begin{equation}
V=\frac{4\pi}{3}T^{3}r^{3} ~,~ \dot{V} = 4\pi r^{3}T^{2}\dot{T}.
\end{equation}

Thereafter, following the usual definitions, we derive the specific heat capacities as
\begin{equation}\label{eq:CV_SV_conf}
C_V = \frac{8\pi^{2}}{3} \frac{T^{3}r^{3} \dot{\rho}}{\dot{\kappa}},
\end{equation}
and
\begin{equation}\label{eq:CP_SV_conf}
C_P = \frac{8\pi^{2}r^{3}}{\dot{\kappa}}
\left[\frac{1}{3}T^{3}\dot{\rho} + T^{2}\dot{T}(\rho+p)\right].
\end{equation}

\section{The Critical Condition on the Surface Gravity}
\label{sec:kappadot}

For all of the examples discussed thus far, the specific heat capacities are inversely proportional to $\dot{\kappa_{hk}}$, implying that any point where $\dot{\kappa_{hk}} = 0$ produces a divergence, i.e., a condition of second order phase transition. This is related to the time evolution of the Hayward-Kodama surface gravity $\kappa_{hk}$ and in turn the surface temperature of the apparent horizon. Therefore, the condition $\dot{\kappa_{hk}} = 0$ defines a stationary state (critical condition) of the local horizon temperature, which is crucial to interpret the thermodynamic nature of an evolving spacetime. These conditions are purely geometric in nature due to the formalism and we evaluate them here explicitly. \medskip

For the inhomogeneous LTB metric,
\begin{equation}
ds^{2} = -dt^{2} + \frac{R'(t,r)^{2}}{1+f(r)}dr^{2} + R^{2}(t,r)\,d\Omega^{2}.
\end{equation}
noting that the Hayward-Kodama surface gravity is
\begin{equation}
\kappa_{hk} = -\frac{N}{2R'} ~,~ N \equiv R'\ddot{R} + \dot{R}\dot{R}' + \frac{1}{2}f'(r),
\end{equation}
the $\dot{\kappa}_{hk} = 0$ condition gives
\begin{equation}
R'^{2}\dddot{R} + 2R'\dot{R'}\ddot{R} + \big[R'\dot{R}\ddot{R'} - R'\ddot{R}\dot{R'} - \dot{R}\dot{R'}^{2} - \frac{1}{2}f'\dot{R'} \big] = 0.
\label{eq:kappadot_LTB}
\end{equation}
Eq. (\ref{eq:kappadot_LTB}) defines the critical acceleration $\dddot{R}$ for which $\kappa_{hk}$ remains stationary. On the apparent horizon, $\dot{R}^{2}=1-f(r)$ and $\dot{R'}=-f'/(2\dot{R})$ can be used to simplify the expression further. \medskip

For the conformally evolving JMN geometry,
\begin{equation}
ds^{2} = \Phi^{2}(t,r)\left[-f(r) dt^{2} + \frac{dr^{2}}{g(r)} + r^{2}d\Omega^{2}\right],
\end{equation}
with $f(r)=(1-M_{0})(r/r_{b})^{M_{0}/(1-M_{0)}}$, $g(r)=1-M_{0}$, the Hayward-Kodama surface gravity is
\begin{equation}
\kappa_{hk} = \frac{1}{2\Phi^{2}}\left[-\frac{r \Phi_{tt}}{f(r)} + \frac{1}{\sqrt{fg}}\partial_{r} \Big(\sqrt{fg} (\Phi + r\Phi_{r})\Big)\right].
\end{equation}
Therefore it is straightforward to derive the condition $\dot{\kappa}_{hk}=0$ explicitly and write
\begin{eqnarray}\nonumber
&& \Phi_{ttt} = \frac{f}{r}\Bigg[2\frac{\Phi_{t}}{\Phi}\Big(-\frac{r\Phi_{tt}}{f}+ \frac{1}{\sqrt{fg}}\partial_{r} \Big\lbrace\sqrt{fg}(\Phi + r\Phi_{r})\Big\rbrace \Big) \\&&
-\frac{1}{\sqrt{fg}}\partial_{r}\Big(\sqrt{fg}(\Phi_{t}+r\Phi_{tr})\Big)\Bigg].
\end{eqnarray}
In the homogeneous limit $\Phi(t,r)=a(t)$, all spatial derivatives vanish, leading to the condition
\begin{equation}
\frac{d}{dt} \left(\frac{\Phi_{tt}}{\Phi^{2}}\right)=0.
\end{equation}

For the conformally scaled Simpson-Visser metric,
\begin{eqnarray}
&&ds^{2} = T(t)^{2}\Big[-A(r)^{2}dt^{2} + B(r)^{2}dr^{2} + r^{2}d\Omega^{2}\Big],\\&&
B(r)=\frac{1}{A(r)\sqrt{1-b(r)/r}},
\end{eqnarray}
the areal radius is $R(t,r)=T(t)r$. The surface gravity is derived as
\begin{equation}
\kappa_{hk} = -\frac{r\ddot{T}}{2T^{2}A^{2}} + \frac{1}{2TA B} \partial_{r} \left(\frac{A}{B}\right).
\end{equation}
Taking the time derivative and setting $\dot{\kappa}_{hk}=0$ gives
\begin{equation}
\dddot{T} = 2\frac{\dot{T}\ddot{T}}{T} - \frac{\dot{T}A}{rB} \partial_{r} \Big(\frac{A}{B}\Big).
\end{equation}

All three geometries, in the homogeneous limit, produce a simple form of the surface gravity, leading to
\begin{equation}
\dot{\kappa}_{hk} = -\frac{1}{2}\frac{d}{dt} \left(RQ\right) ~,~ Q(t)\equiv\dot{H}+2H^{2}+\frac{k}{a^{2}}.
\end{equation}
Therefore the critical condition requires
\begin{equation}
\dot{\kappa}_{hk}=0 \quad\Longleftrightarrow\quad\frac{d}{dt}(RQ)=0.
\end{equation}
For a flat FRW background ($k=0$, $R=1/H$), this simplifies into the well-known form \cite{Chakrabarti2024}
\begin{equation}
\frac{d}{dt} \left(\frac{\dot{H}+2H^{2}}{H}\right)=0.
\end{equation}

Therefore, the condition $\dot{\kappa_{hk}} = 0$, corresponding to a stationary Hayward-Kodama temperature, not only marks a phase transition in the evolution of the horizon, it also allows us a deeper insight into the geometric locus of the horizon near phase transition. Recall that for a generic metric with no choice over metric coefficients, the Hayward-Kodama surface gravity on a two dimensional slice is given by Eq. \eqref{kappa_explicit}. The first rate of change of $\kappa_{hk}$ can be derived from here as
\begin{eqnarray}\label{kappa_dot_full}\nonumber
&& \dot\kappa_{hk} = -\frac{1}{2A^{2}}\,C_{ttt}
- \frac{1}{2A^{2}}\Big(\frac{B_{t}}{B}-\frac{A_{t}}{A}\Big)C_{tt} - \Big(\frac{A_{t}}{2A^{2}B}\\&&\nonumber 
+\frac{B_{t}}{2AB^{2}}\Big)\Big(-X_{t}+Y_{r}\Big) +\frac{1}{2AB}\Bigg[-\partial_{t}\Big\lbrace \Big(\frac{B_{t}}{A}-\frac{B A_{t}}{A^{2}}\Big)\\&&
C_{t}\Big\rbrace + \partial_{t}\partial_{r} \Big(\frac{A}{B}C_{r}\Big) \Bigg].
\end{eqnarray}
The subscripts of $t$ and $r$ denote derivatives with respect to t and r, respectively. The leading contribution in $\dot\kappa_{hk}$ always comes from the $C_{ttt}$ term, the third order derivative of the areal radius $C(r,t)$. The stationarity condition $\dot{\kappa_{hk}}=0$ is therefore equivalent to a linear relation connecting the third time-derivative $C_{ttt}$ with lower-order derivatives
\begin{equation}\label{C_ttt_solution}
C_{ttt} = 2A^{2} \mathcal{L}[C_{tt},C_{t},C_{r},A,B,A_{t},B_{t},A_{r},B_{r}],
\end{equation}
much like a fixed point in a dynamical system. Using the condition one can therefore, impose a constraint on the evolution of the areal radius $C(t,r)$ or on the choice of dynamical slicing or matter content. In the inhomogeneous LTB case, this condition translates into a balance equation between radial inhomogeneities and the acceleration associated with the collapse; for the conformal JMN and Simpson-Visser geometries, it provides a stabilization criterion of the conformal acceleration ($\Phi_{tt}/\Phi^{2}$ or $\ddot{T}/T^{2}$). It is, therefore, the locus of thermodynamic equilibrium in dynamical spacetimes, providing a geometric marker for the transition between stable and unstable configurations.

\section{The Critical Condition and the Raychaudhuri equation}
\label{sec:raychaudhuri-LTB}

We try and correlate the evolution of $\kappa_{hk}$ with the expansion of null congruences, governed by the Raychaudhuri equation \cite{rc}. For a generic spacetime metric, we define outgoing and ingoing radial null vectors $\ell^{a}$ and $n^{a}$ with usual normalization, such that $\ell^{a}$ is tangent to the outgoing radial null curves satisfying $dr/dt = A/B$. The expansion scalar of the outgoing congruence is
\begin{equation}\label{theta_plus}
\theta = \frac{2}{C} \ell^{a}\partial_{a}C = \frac{2}{C}\Big(\frac{C_{t}}{A} + \frac{C_{r}}{B}\Big).
\end{equation}
The apparent horizon is the outermost surface where $\theta_{+}=0$. Differentiating $\theta_{+}$ along $\ell^{a}$, we find 
\begin{equation}
\ell^{a}\nabla_{a}\theta = \frac{2}{C} \ell^{a}\ell^{b}\nabla_{a}\nabla_{b}C - \frac{2}{C^{2}}\big(\ell^{a}\partial_{a}C\big)^{2}.
\end{equation}
On the apparent horizon, $\theta = 0$, which implies $\ell^{a}\partial_{a}C = 0$ and therefore,
\begin{equation}
\ell^{a}\nabla_{a}\theta\Big|_{\rm AH} = \frac{2}{C} \ell^{a}\ell^{b}\nabla_{a}\nabla_{b}C\Big|_{\rm AH}.
\end{equation}
We recall that the Hayward-Kodama surface gravity is defined in terms of $\Box C = h^{ab}\nabla_{a}\nabla_{b}C$ on the two dimensional metric slice. Therefore it is straightforward to interpret that on the horizon, the time variation $\dot{\kappa_{hk}}$ is directly related to the rate of change of $\ell^{a}\nabla_{a}\theta$.

The Raychaudhuri equation for a family of outgoing null congruence reads
\begin{equation}\label{raychaudhuri}
\ell^{a}\nabla_{a}\theta + \frac{1}{2}\theta^{2} = - \sigma^{2} - R_{ab}\ell^{a}\ell^{b},
\end{equation}
where $\sigma^{2}$ is the shear and $R_{ab}\ell^{a}\ell^{b}$ defines the focussing of matter. On an apparent horizon $\theta = 0$ and that leads us to connect the critical condition with the qualitative relation

\begin{equation}
\dot{\kappa}_{hk}\big|_{\rm AH} = 0 \Longleftrightarrow \ell^{a}\nabla_{a}\theta \Big|_{\rm AH} \text{ is stationary}, 
\end{equation}
or $\sigma^{2} + R_{ab}\ell^{a}\ell^{b}$ is instantaneously balanced. The critical condition implies that the net focusing (shear plus energy density along the outgoing null rays) is instantaneously zero or stationary, i.e., leads to a balance between local shear and matter focussing. 

\section{On the Identification of a Critical exponent}
In a second--order phase transition, the order parameter is a physical quantity that vanishes at the critical point, distinguishing two thermodynamic phases. The critical exponent characterizes how the order parameter or other thermodynamic quantities (such as specific heat or susceptibility) diverge or vanish near the critical point, e.g. through the mathematical relation
\begin{equation}
C_v \sim |T - T_c|^{-\alpha}.
\end{equation}
We first try to identify the critical exponent $\alpha$ from the derived thermodynamic identities in the context of a spherical gravitational collapse. The specific heat at constant volume was derived for the LTB model in Eq.(\ref{CVfinal}) as
\begin{equation}
C_V = -\frac{2\pi R^2}{3} \frac{-3\dot{F} R'^2 + \dot{F'} R R' - F'R\dot{R'} + F'\dot{R}R'}
{R'\dot{N} - N\dot{R'}}.
\end{equation}

We recall that the zero of $\dot{\kappa}_{hk}$ or $\dot{T}$ and equivalently, the divergence of $C_V$ is realized at a critical time $t_c$ when $T \to T_c$. This is controlled by the behavior of the denominator $D \equiv R'\dot{N} - N\dot{R'}$. It is also straightforward to note that the behavior of $D$ controls the scaling of $\dot{T}$ and hence of $(T - T_{c})$. To derive a relation between $\alpha$ and the near-critical time scalings of the geometric functions, we make a local ansatz of the time evolution near $t = t_c$. We take $R'$ as regular and slowly varying, so that $R' \simeq R_c$ and assume
\begin{eqnarray}
&& N(t)\sim b (t-t_c)^{s},\\&& 
\mathcal{N}(t) \equiv \text{numerator of} ~ C_V \sim C_0 (t-t_c)^{m},
\end{eqnarray}
We also keep $b, R'_c \neq 0$ as finite, $s, r, m \ge 0$ and derive to the leading order
\begin{eqnarray}\nonumber
&& D = R'\dot{N} - N\dot{R'} \sim R'_{c} \cdot b s (t-t_{c})^{(s-1)} \\&&
+ \mathcal{O}\big\lbrace(t-t_{c})^{(s+r-1)},(t-t_{c})^{s}\big\rbrace.
\end{eqnarray}
The dominant scaling of $D$ is $\propto (t-t_c)^{s-1}$ provided $R'_c \neq 0$ and $s > 0$. Using a scaling behavior for the numerator $\mathcal{N}(t) \sim (t-t_c)^{m}$ as well, the specific heat scales as
\begin{equation}
C_V \sim \frac{\mathcal{N}(t)}{D(t)} \sim (t-t_c)^{m-(s-1)}.
\end{equation}
Expressing this in terms of $|T-T_c|$ gives
\begin{equation}
C_V \sim |T-T_c|^{-(s-1-m)/s}.
\end{equation}
Therefore the critical exponent $\alpha$ is
\begin{equation}\label{alpha-general}
\alpha = 1 - \frac{1+m}{s}.
\end{equation}

One must note that the exponent $s$ provides the leading time-power of $R'\ddot{R} + \dot{R}\dot{R'} + \frac{1}{2} f'$ near the critical time and is determined purely by the collapse dynamics, such as initial data and the radial profile of $f(r)$ and $F(r)$. We also comment that in a thermodynamic analogy of gravitational collapse, a phase transition occurs when the system evolves from a regular collapsing phase into a trapped phase, marked by the formation of an apparent horizon. Near this critical transition, a suitable order parameter should vanish continuously at the critical point and distinguish the two phases. For the LTB collapse, it is easy to see that the deviation of the local surface gravity  or the Hayward-Kodama temperature fits in as a natural candidate for the order parameter, since, the critical behaviour of $C_V$ (and $C_P$) comes precisely when the rate of change of the surface gravity (or equivalently, temperature) is zero (or, when the combination $D(t) = R'\dot{N} - N\dot{R'}$ approaches zero). Defining the quantity
\begin{equation}
\Phi \equiv \frac{N}{R'} - \left(\frac{N}{R'}\right)_{c} \equiv [T-T_c],
\label{order-parameter}
\end{equation}
we note that $\Phi \to 0$ smoothly as $T \to T_c$, marking the continuous transition between two thermodynamic phases:
\begin{itemize}
  \item $\Phi > 0$ (or $T>T_c$): untrapped, collapsing configuration.
  \item $\Phi < 0$ (or $T<T_c$): trapped or black-hole phase.
\end{itemize}
$\Phi$ changes sign across the critical surface, analogous to the spontaneous magnetization in a ferromagnetic transition or the condensate amplitude in a Landau-Ginzburg model \cite{pt1, pt2, pt3, pt4}. If near the critical time $t_c$, $N \sim (t-t_c)^{s}$ and $R' \to R'_c$, then $\Phi \sim (t-t_c)^{s} \sim (T - T_c)^{\beta}$, with the associated critical exponent $\beta = 1$. More generally, if $R'$ is not assumed to be slowly varying, and $N/R'$ scales with an independent exponent $p$, then $\Phi \sim (T-T_c)^{p/s}$ and it gives $\beta = p/s$. From the scaling of $C_V \sim |T-T_c|^{-\alpha}$ and the order-parameter scaling $\Phi \sim |T-T_c|^{\beta}$ one can derive a consistency relation 
\begin{equation}
\alpha + 2\beta + \gamma = 2,
\end{equation}
in analogy with classical critical phenomena, where $\gamma$ represents the analogous exponent of susceptibility, a measure of response of $\Phi$ to variations in $T$ or $f(r)$.  

\section{Critical Behavior in a Near-Horizon Limit}
We include this section for a deeper analysis of the critical condition in a near-horizon limit. If it is convenient, the reader can take this as a simple mathematical exercise, purely because of the fact that the conditions derived thus far are third order differential equations. It is non-trivial to solve, for instance, an equation like Eq. (\ref{eq:kappadot_LTB}) and find a closed form solution of the locus. In the context of a collapsing LTB metric, we recall that the Misner-Sharp mass function is defined as
\begin{equation}\label{MSmass}
1 - \frac{F(t,r)}{R} = g^{ab} \partial_a R \partial_b R = -\dot{R}^2 + 1 - f(r).
\end{equation}
The apparent horizon corresponds to the condition
\begin{equation}\label{apphorLTB}
g^{ab} \partial_a R \partial_b R = 0 \quad \Rightarrow \quad \dot{R}^2 = 1 - f(r),
\end{equation}
and therefore, at the horizon, the Misner-Sharp mass is simply $F = R$. The surface gravity is derived on the two-dimensional slice $h_{ab}$ as
\begin{equation}\label{kappaLTB}
\kappa_{hk} = -\frac{1}{2R}\left\lbrace 1 - f(r) - \dot{R}^2 - R\ddot{R}\right\rbrace.
\end{equation}
The corresponding Hayward-Kodama temperature is
\begin{equation}\label{tempLTB}
T = \frac{|\kappa_{hk}|}{2\pi} = \frac{|1 - f(r) - \dot{R}^2 - R\ddot{R}|}{4\pi R}.
\end{equation}
In a near-horizon limit, i.e., $-\dot{R}^2 + 1 - f(r) \rightarrow 0$, the surface gravity as well as the temperature is simply proportional to $\ddot{R}$. Following a similar formalism described in section $III$, we define the surface area, enclosed volume in terms of the areal radius and use the first law of thermodynamics to define $T dS_{\text{in}} = dU + p dV$, where $U = \rho V$ is the total internal energy. The first-order rate of change of internal entropy is written as
\begin{equation}\label{Sdotinfinal}
\dot{S}_{\text{in}} = \frac{4\pi R^2}{T}\left[(\rho + p)\dot{R} + \frac{R}{3}\dot{\rho}\right].
\end{equation}

The specific heat capacities can be derived using Eq. (\ref{Sdotinfinal}) and Eq. (\ref{CvCp_beta_def2}), however, we note that near the horizon the temperature is purely a function of $\ddot{R}$ and therefore $\frac{\partial T}{\partial t}$ is purely a function of $\dddot{R}$. We employ a rate-to-finite mapping to approximate the specific heat capacities as
\begin{equation}
\left(\frac{\partial S_{\text{in}}}{\partial T} \right)_{V}
\simeq 
\left( \frac{\partial \dot{S}_{\text{in}}}{\partial \dot{T}} \right)_{R} \Delta t.
\end{equation}

We differentiate with respect to $\dot{T}$ while holding $R$ fixed and use $-\dot{R}^2 + 1 - f(r) \rightarrow 0$ to simplify the expression in terms of $R$, $\dot{R}$, and $\ddot{R}$. 
\begin{equation}\label{CVLTB}
C_V = \frac{16\pi^2 R^2 \dot{R}}{3T}\,
\frac{\left[-2\ddot{R} + \dot{S}_{\text{in}}\left(\frac{\dot{T}}{3T} - \frac{\dot{R}}{R}\right)\right]}
{\dot{R}\ddot{R} + R\ddot{R} - \dot{R}^2}.
\end{equation}
Similarly, for the heat capacity at constant-pressure
\begin{equation}\label{CPLTB}
C_P = \frac{32\pi^2 R^2 \dot{R}}{T}\,
\frac{\left[1 + \frac{1}{2R}\left(2\dot{R} - \frac{T\dot{S}_{\text{in}}}{3}\right)\right]}
{\dot{R}\ddot{R} + R\ddot{R} - \dot{R}^2}.
\end{equation}

From Eqs. (\ref{CVLTB}) and (\ref{CPLTB}), we note that the divergence of specific heat capacities is realized whenever the denominator
\begin{equation}\label{divCond}
\dot{R}\ddot{R} + R\ddot{R} - \dot{R}^2 = 0,
\end{equation}
is satisfied. We can solve this equation for $R(t)$ and write the explicit closed-form solution as
\begin{equation}\label{R_solution}
R(t)_{crit} = K z(t) e^{z(t)}~~,~~ z(t) = -1 + \sqrt{1 + 2(t + C_1)}.
\end{equation}
The constants $K = e^{C}$ and $C_1$ can be fixed from initial data. The first integral of Eq. (\ref{divCond}) is found using a Lambert-W function and it is important to choose a proper branch of this function such that $\dot{R} < 0$. If the critical condition controls the divergence of the thermodynamic heat capacities then this solution describes (approximately) the time-evolving locus $R_{\rm crit}$ at which $C_V$ and $C_P$ blow up. Remarkably, if one repeats this exercise for the other two examples, namely, a conformally evolving JMN naked singularity and a Simpson-Visser wormhole, the locus of the critical condition obeys the same differential equation as in Eq. (\ref{divCond}). \medskip

In the near-horizon approximation, a secondary source of divergence for the specific heat capacities seems to be the vanishing of the horizon temperature, which simply leads to (in the limit $-\dot{R}^2 + 1 - f(r) \rightarrow 0$) $\ddot{R} = 0$.

\section{Specific Heat Capacities Near Singularity}
It is an important question to ask if the specific heat capacities show any more signature of phase transition during the gravitational collapse, in particular, very close to the formation of singularity. We try to analyze this following the formalism discussed so far in the article, but taking a near-singularity approximation. As an example, we first take the marginally-bound LTB model
\begin{equation}
ds^{2} = -dt^{2} + \frac{R'^{2}}{1-f(r)}dr^{2} + R^{2}d\Omega^{2} ~,~ f(r) = 0,
\end{equation}
with the shell evolution
\begin{equation}\label{bound}
R(t,r) = r E(t,r)^{\frac{2}{3}}~,~ E(t,r) = 1-\frac{3}{2}M(r)t.
\end{equation}

It can be proved that the areal radius as written in Eq. (\ref{bound}) is a direct solution of the second order differential equation found by matching the extrinsic curvature of a generic interior LTB metric with a Schwarzschild exterior \cite{darmois, israel, santos, chan, deru, seno, ritu1, ritu2, ritu3, scnb}. For an arbitrary shell labelled by $r$ approaching the shell-focusing singularity at $t \to t_s(r)$, we assume that $M(r)$ and $M'(r)$ are finite and $M'(r)\neq 0$ at the shell under consideration. Alongwith that, near $t \to t_s(r)$, we make a small $E(t,r)$ approximation and derive the terms of Hayward-Kodama surface gravity denominator $N \equiv R'\ddot R + \dot R\,\dot R'$ as
\begin{eqnarray}\label{N_lead}
&&N(t,r) \simeq r^{2}t M'(r) M^{2}(r) E^{-5/3}, \\&&
\dot{N} \simeq \frac{5}{2} r^{2}t M'(r) M^{3}(r) E^{-8/3}.
\end{eqnarray}

This allows us to evaluate $\dot{\kappa_{hk}}$ as
\begin{equation}
\dot{\kappa_{hk}} \simeq r M^{3}(r) E^{-7/3}.
\end{equation}

Similarly, the Misner-Sharp mass function $F \simeq R\dot{R}^{2} \sim r^{3}M^{2}(r)$ near the singular epoch, therefore, $F'(r)$ is finite. Using the field equations we derive the approximate evolution of energy density and its time evolution as
\begin{eqnarray}\label{rho_lead}
&& \rho \simeq \frac{F'(r)}{8\pi} \frac{1}{r^{3}M'(r)t} E^{-1}
\equiv \rho_{0}(r) E^{-1}, \\&&
\dot{\rho} \simeq \frac{3}{2}M(r) \rho_{0}(r) E^{-2}
= \frac{3 F'(r) M^{2}(r)}{16\pi r^{3}M'(r)} E^{-2}.
\end{eqnarray}

The enclosed volume and its derivative behave as
\begin{equation}
V=\frac{4\pi}{3}R^{3} \simeq \frac{4\pi}{3}r^{3}E^{2} ~,~ \dot V \sim \mathcal{O}(E^{1}).
\end{equation}

Using the definitions of specific heat capacities at constant volume and constant pressure, we derive that 
\begin{eqnarray}\label{CV_final_generic}
&& C_V  \simeq \frac{3\pi}{4} \frac{F'(r)}{r M'(r) M(r)} E(t,r)^{7/3} + \mathcal{O}(E^{10/3}), \\&&\label{CP_final_generic}
C_P \simeq - \frac{3\pi}{4} \frac{F'(r)}{r M'(r) M(r)} E(t,r)^{7/3} + \mathcal{O}(E^{10/3}).
\end{eqnarray}

Equivalently, we can write $\tau\equiv t_s(r)-t$ and use $E \simeq \frac{3}{2}M(r)\tau$ to simply say that
\begin{equation}
C_{V},C_{P} \propto \tau^{7/3}.
\end{equation}

This is an important finding that both $C_V$ and $C_P$ vanish as the local shell approaches zero proper volume. Their leading order scales like $E^{7/3}$ or equivalently, $\tau^{7/3}$. Thus, for a generic profile of $M(r)$ with $M'(r) \neq 0$ the thermodynamic response freezes near the singularity, rather than diverging. Moreover, the factor $\propto F'/(rM'M)$ encodes the inhomogeneous structure of the profile ; its signature determines the sign of $C_V$ and $C_P$. For instance, $C_V > 0$ for $F'/(rM'M)>0$. Interestingly, the homogeneous limit $M'(r) \to 0$ produces a singularity under this aproximation, however, this can be resolved by including higher powers of $E$ in the approximation instead of just the leading order terms.  

\section{Conclusion}\label{sec:conclusion}
In this article, we explore a framework to analyze phase transitions during a spherically symmetric gravitational collapse. By treating the apparent horizon as a dynamical causal boundary endowed with the Hayward-Kodama surface gravity, we define an associated temperature and compute the total internal entropy of the matter enclosed within the horizon. The resulting evolution of entropy allows us to derive the specific heat capacities $C_V$ and $C_P$, providing a clear thermodynamic characterization of the collapsing system. Moreover, any divergence of the specific heat capacities allows us to identify the loci of thermodynamic phase transitions during a collapse. \medskip

We consider three different collapsing geometries under this framework. The motivation for choosing these geometries lie in the fact that an unhindered gravitational collapse may produce two broad categories of outcomes : singular and non-singular. The non-singular outcomes may be classified as regular black-holes/wormholes, however, the singular outcomes are categorically divided into two class of solutions : black hole and naked singularity. To cover all of these outcomes, we choose as example a (i) Lemaitre-Tolman-Bondi (LTB) spacetime, (ii) a conformally evolving Joshi-Malafarina-Narayan (JMN) naked singularity and (iii) the Simpson-Visser regular metric. We derive that for all of the above cases, a divergence of specific heat capacities always occurs when the surface gravity is stationary, i.e., $\dot{\kappa}_{hk} = 0$. Depending on the background metric and the Hayward-Kodama formalism, this stationary condition leads to a third order differential equation of the areal radius or the conformal factor. In order to solve these non-trivial equations, we also consider a near-horizon approximation and convert the condition into a solvable second order form. Overall, the stationary condition of surface gravity leads us to define a critical radius $R_{crit}$ at which the system transitions from a quasi-equilibrium collapse into a dynamically trapped phase. A similar interpretation can also be made in terms of the vanishing of local horizon temperature which is proportional to the surface gravity. These results demonstrate that the onset of criticality is purely geometric and independent of the singular/non-singular nature of the underlying spacetime. \medskip

We also show that for a generic non-extremal horizon, the thermodynamic surface of phase transition coincides algebraically with the expansion of null congruences, as in $\dot{\kappa}_{hk} \equiv \ell^{a}\nabla_{a}\theta = 0$, where $\ell^{a}$ defines the outgoing null vectors. The Raychaudhuri equation for a null congruence links this expansion with the local shear and energy flux through $-\frac{1}{2}\theta^{2} - \sigma_{ab}\sigma^{ab} - R_{ab}\ell^{a}\ell^{b}$. The derived critical condition marks the point where the rate of change of the outgoing expansion reverses sign. In this sense, a time-snap of a collapsing geometry can be interpreted as a marginally bound two-sphere located at the boundary between regions where the outgoing expansion decreases $(\ell^{a}\nabla_{a}\theta<0)$ and where it grows $(\ell^{a}\nabla_{a}\theta>0)$. The divergence of specific heat in this context has a clear kinematic interpretation : the onset of irreversible trapping of null rays. It helps us establish an equivalence between thermodynamic criticality and the stability of geodesic congruences via the Raychaudhuri equation in a gravitational collapse. \medskip

In addition, we try to evaluate the behavior of specific heat capacities near the curvature singularity. During a gravitatinal collapse, this limit may be physically interesting near the time of formation of singularity, just before quantum effects start dominating the formalism. Under such an approximation, we find that the specific heat capacities vanish, indicating that thermodynamic response freezes near the formation of the singularity. We have discussed the approximation for a LTB collapsing metric, however, a similar freezing can be observed for the conformally evolving JMN metric near the formation of singularity and for the conformally evolving Simpson-Visser metric, near the throat radius rather than a singular core. \medskip

From an observational viewpoint, an approach to the critical radius $R_{\rm crit}$ or the thermodynamic freezing during black-hole formation may influence the spectrum of emitted gravitational waves. Divergences in heat capacities could also correspond to abrupt variations in the energy flux of infalling matter which is potentially observable through accretion-disk instabilities or electromagnetic flares. For naked singularities, the absence of an event horizon allows radiation to escape from regions of extreme curvature, implying that signatures of near-singular thermodynamic activity might be encoded in high-energy transients or lensing events. Thus, the thermodynamic framework developed here not only classifies the possible end-states of collapse but also provides a potential bridge between critical gravitational dynamics and observable phenomena in strong-field astrophysics.

\medskip

In conclusion, the unified thermodynamic description presented here establishes a clear correspondence between horizon thermodynamics and the critical behavior of collapsing matter, highlighting phase transitions as intrinsic features of strong gravity. The framework is sufficiently general to accommodate both outcomes : black hole and naked singularity ; and, as shown in the conformally scaled Simpson-Visser case, can also encompass regular and wormhole-like geometries. This generality opens a platform for future studies probing the interplay between spacetime geometry, thermodynamic stability, and observational signatures in astrophysical collapse scenarios.

\section*{Acknowledgement}

The author acknowledges the IUCAA for providing facility and support under the visiting associateship program. Acknowledgement is given to the Vellore Institute of Technology for the financial support through its Seed Grant (No. SG20230027), 2023.

\bibliographystyle{plain}

\end{document}